\documentstyle[epsf,prl,preprint,tighten,aps]{revtex}
\begin{document}
\preprint{G\"oteborg ITP 97-08}
\draft
\title{Correlation Functions and Coulomb Blockade of Interacting 
Fermions at Finite Temperature and Size}
\author{Sebastian Eggert$^a$, Ann E. Mattsson$^{a}$, and Jari M. Kinaret$^b$}
\address{
$^a$Institute of Theoretical Physics and $^b$Department of Applied Physics\\
Chalmers University of Technology and G\"oteborg University,
S-412 96 G\"oteborg, Sweden}
\maketitle
\begin{abstract}
\widetext\leftskip=0.10753\textwidth \rightskip\leftskip
We present explicit expressions for the correlation functions
of interacting fermions in one dimension which are valid for
arbitrary system sizes and temperatures.  The result
applies to a number of very different strongly correlated systems, including
mesoscopic quantum wires, quantum Hall edges, 
spin chains and quasi-one-dimensional 
metals.  It is for example possible to calculate Coulomb blockade oscillations
from our expression and determine their dependence on interaction strength
and temperature.  Numerical simulations show excellent agreement 
with the analytical results. 
\end{abstract}
Recently, there has been great interest in strongly correlated systems of 
all kinds, which is spurred by high temperature superconductivity,
progress in mesoscopic physics, quantum Hall systems, and newly
available materials which accurately display the characteristics of 
one-dimensional metals or spin chains.  In the case of quasi-one-dimensional
electron systems much progress can be made with 
the Luttinger liquid formalism\cite{Voit} which is able to describe
virtually any type of local interaction and can be solved by use of 
bosonization techniques\cite{Haldane,Stone}.  
Using these methods, it is well known how to 
calculate correlation functions in the infinite length or zero temperature 
limit and therefore determine the spectral properties\cite{meden} 
of a number of systems, like mesoscopic quantum wires, 
quantum Hall bars\cite{Wen}, quasi one-dimensional metals and 
spin chains\cite{luther,schulz}
even in the presence of boundaries\cite{eggert,boundary}.
We now present explicit expressions for the correlation functions 
which are valid at arbitrary temperature, system size, and distances (as long
as the lattice spacing is relatively small).  These expressions
can be used to calculate Coulomb blockade oscillations in mesoscopic 
systems as a function of the interaction strength and of temperature.  
Monte Carlo simulations of one of the simplest systems, namely 
interacting spinless fermions on a one-dimensional lattice, show excellent 
agreement with our result.

As our model Hamiltonian we consider one-dimensional interacting fermions
in the continuum limit
\begin{equation}
\label{ham}
{H} \  = \ \int_0^{\ell} dx \ \ \left[    v_F\ 
  \psi^\dagger_{L} i {d \over dx} \psi^{}_{L}
 - v_F \ \psi^\dagger_{R} i {d \over dx} \psi^{}_{R} 
  \ + \    g_1 (J_{L}   J_{L} \ + \    J_{R}   J_{R}) 
  \ + \    g_2  J_{L}   J_{R} \right]
\end{equation}
Here $J_{L/R}
 \equiv \ : \! \!\psi_{L/R}^\dagger\psi_{L/R}^{}\! \!:$ are the
chiral fermion currents of the left- and right-moving components
of the fermion field $\Psi(x)$, expanded about the
Fermi points $\pm k_F$, $\Psi(x) = 
e^{-ik_Fx}\psi_{L}(x) + e^{ik_Fx}\psi_{R}(x).$
This expansion is valid as long as the lattice spacing $a$ is small
compared to the inverse temperature $v \beta$ and the length scales 
we want to consider.   In this limit the Hamiltonian (\ref{ham}) can
describe almost any one-dimensional fermion system with short range 
interactions by choosing appropriate coupling constants $g_1$ and $g_2$.
The umklapp process $e^{i 4 k_F x}\   \psi^\dagger_{L}  \psi^{}_{R}
 \psi^\dagger_{L}  \psi^{}_{R} + h.c.$ can also be included, but it
only contributes if the system is close to half-filling $k_F=\pi/2a$ 
and then it can often be absorbed by renormalizing the coupling 
constant $g_2$\cite{Voit}.  

The fermion correlation functions of systems with the Hamiltonian of equation
(\ref{ham}) are known to obey simple power-laws in the zero-temperature and
infinite length limit, with exponents determined by the 
interaction\cite{Voit}.
At non-zero temperatures the correlation functions are 
exponentially damped and are described by powers of the hyperbolic sine
in the infinite length limit.
For finite length $\ell$ and periodic boundary conditions, on the other hand, 
the correlation functions are periodic and at zero temperature 
are described by powers of the sine instead.  The correlation functions
are essential to understand the spectral characteristics
and other properties of experimental systems, and it is often necessary
to consider both finite size and finite temperature. We  will now
derive explicit expressions for this case and show the crossover
between the two limits.   

To establish notation we will review shortly 
the bosonization formalism\cite{Stone} for the Hamiltonian (\ref{ham}),
which is given by the relations 
\begin{equation}
J_{L,R} = \frac{1}{4 \pi} \partial_x (\phi \pm \tilde \phi), \ \ \ \ 
\psi_{L,R} \propto e^{-i \sqrt{\pi} (\tilde \phi \pm {\phi}) } \label{bose}
\end{equation}
where $\tilde{\phi}$ is the dual field
to the boson $\phi$ with the finite length mode expansion
\begin{eqnarray}
\phi(x,t) & = &
\phi_0 \, + \,  \Pi{vt \over \ell} + Q{x\over \ell}
 \, + \, \sum_{n =1}^{\infty} \frac{1}{\sqrt{4 \pi n}}
\left[ e^{-\frac{2\pi i}{\ell} n(vt + x)} a_n^L \,+\,  e^{-\frac{2 \pi
i}{\ell} n(vt - x)} a_n^R + \, h.c. \right] \nonumber \\
\tilde \phi (x, t) &  =  & \tilde \phi_0 \, + \, Q{vt \over \ell} \, +\, 
\Pi {x\over \ell}
 \, + \, \sum_{n =1}^{\infty} \frac{1}{\sqrt{4 \pi n}} \left[ e^{-\frac{2
\pi i}{\ell} n(vt + x)} a_n^L \,-\,  e^{-\frac{2 \pi
i}{\ell} n(v t - x)} a_n^R + \, h.c. \right]. \label{mode}
\end{eqnarray}
Here $\Pi$ and $Q$ are canonically conjugate to the zero modes
$\phi_0$ and $\tilde \phi_0$, respectively, and the
time variable implicitly carries a small ultraviolet cutoff $t - i \alpha$.

The $g_1$ interaction can be absorbed by redefining
the velocity $v = v_F + g_1/2 \pi$, and the Hamiltonian density becomes
\begin{equation} {\cal H}  \ =  \    \frac{v}{2}
\left[({\partial_x \phi})^2 + ({\partial_x \tilde \phi})^2\right]
\ + \ \frac{g_2}{4 \pi}
\left[({\partial_x \phi})^2 - ({\partial_x \tilde \phi})^2\right].
\end{equation}
The Hamiltonian
is then solved by a simple rescaling of the boson, 
\begin{equation}
\phi \to K \phi, \ \  \tilde \phi \to \tilde \phi/K, \label{rescale}
\end{equation}
where $K= 1 - g_2/4 \pi v$ to lowest order in the coupling constant (i.e.
$K=1$ for a non-interacting system and $K < 1$ for repulsive interactions).

The eigenvalues of the operators $\Pi$ and $Q$ are quantized by using the 
relation for periodic boundary conditions 
$\psi_{R,L}(0,t) = e^{\pm i k_F \ell} \psi_{R,L}(\ell,t)$ together with
equations (\ref{bose}) and (\ref{mode}). After a careful consideration
of the commutation relations we find
\begin{equation}
Q \ = \ \sqrt{\pi} (n-n_0)/K,  \ \ \  \Pi \ = \ \sqrt{\pi} K m, \label{quanta}
\end{equation}
where $n$ and $m$ are either both even or both odd integers and 
$n_0 = \frac{k_F \ell}{\pi} + 1$ which can be defined modulo 2.  The 
quantum number $n$ represents the particle number 
and $m$ is a measure of the current in the ring.  
If a  magnetic field is applied, the ground state may have
quantum number $m\ne 0$  and the system carries a
persistent current\cite{loss,kinaret}.

As a simple example of our calculation at finite temperature and size,
we now want to consider the chiral Green's function 
\begin{equation}
G(x,t) \equiv \left<\psi^{}_L (x,t) \psi^\dagger_L (0,0)\right>
\end{equation}
but the generalization to more complicated correlation functions is 
straightforward as we will see later.
Using equations (\ref{bose}) and (\ref{rescale}) and applying the
Baker-Hausdorf formula as well as the cumulant theorem for bosonic modes,
we obtain 
\begin{eqnarray}
\label{GF}
G(x,t) & \propto &\left< \exp -i \sqrt{\pi}
\left[K\phi(x,t)+\tilde\phi(x,t)/K\right]
\exp  i\sqrt{\pi}
\left[K \phi(0,0) + \tilde \phi(0,0)/K\right]
\right>   \\
& = & e^{-i \frac{\pi}{\ell} [x + \frac{K^2 + 1/K^2}{2} vt]} \
B(x,t) \ \exp\left[\pi (K+ K^{-1})^2 G_L(x,t)\right]
\exp\left[\pi (K- K^{-1})^2 G_R(x,t)\right] \nonumber
\end{eqnarray}
where $B(x,t) \equiv\left<\exp -i \sqrt{\pi} 
\left(Q \frac{K x + v t/K}{\ell} +
\Pi \frac{K v t + x/K}{\ell}\right) \right>$ is the contribution 
from the operators $\Pi$ and $Q$, and   
$G_{L,R} \equiv \left<\phi_{L,R}(x,t) \phi_{L,R}(0,0) - 
\phi_{L,R}^2(0,0)\right>$ are the chiral boson Green's functions for the
left- and right-moving modes 
$\phi_{L,R} = [\phi \pm \tilde \phi - (Q \pm \Pi)\frac{x \pm vt}{\ell}]/2$,
respectively. Using the mode expansion (\ref{mode}) the chiral boson
Green's function can be expressed as
\begin{equation}
G_L(x,t) 
\ = \   \frac{1}{4 \pi} \sum_{n=1}^{\infty} \frac{1}{n}
\left[  (e^{-i 2 \pi n  \frac{x+v t}{\ell}} -1) \left<a_n^L 
{a_n^L}^\dagger\right> + (e^{i 2 \pi n  \frac{x+v t}{\ell}} -1) 
\left<{a_n^L}^\dagger a_n^L \right>\right].
\end{equation} 
Since we know the energy levels 
$E_n = n\frac{2 \pi v}{\ell}$ of the free boson modes,
it is useful to write the Bose-Einstein distribution as a sum 
over occupation numbers
 $\left<a_n^\dagger a_n\right> = 
 \frac{1}{e^{\beta E_n}-1} = \sum_{k=1}^\infty 
e^{-\frac{2 \pi v}{\ell} \beta k n}$.
After exchanging the order of the summations over $k$ and $n$ we can
use the identity $\sum_{n=1}^\infty \frac{x^n}{n} = - \ln(1-x)$ to obtain
\begin{equation}
G_L(x,t) \ = \ - \frac{1}{4 \pi} \ln \left[ 
\frac{1- e^{- 2 \pi i (x+vt)/\ell}}{\pi \alpha/\ell} 
\prod_{k=1}^\infty
\frac{1-e^{-2 \pi i (x + vt -i v \beta k)/\ell}}{1 - e^{-2 \pi v \beta k/\ell}}
\frac{1-e^{2 \pi i (x + vt +i v \beta k)/\ell}}{1 - e^{-2 \pi v \beta k/\ell}}
\right] \end{equation}
Using the elliptic theta function of the first kind $\theta_1(z,q)$\cite{GR},
this result can be written more compactly as
\begin{equation}
G_L(x,t)  = i\frac{x+vt}{4 \ell}  - \frac{1}{4 \pi} \ln \frac{\theta_1(
\pi\frac{x+vt}{\ell}, e^{-\pi v \beta/\ell})}{\theta_1(
-i\frac{\pi \alpha}{\ell},  e^{-\pi v \beta/\ell})}
\end{equation}
and likewise for $G_R(x,t) = G_L(-x,t)$.

The cumulant theorem does not apply to the contribution $B(x,t)$
 from the operators $\Pi$ and $Q$,
but we can use the energy spectrum $E = \frac{v}{2\ell}(Q^2 + \Pi^2)$ and
sum over all eigenvalues in equation (\ref{quanta}). 
The result can again be expressed
in terms of the elliptic theta functions\cite{GR}
\begin{equation} \label{zeromodes}
B(x,t)  
 =  e^{ in_0 x_Q}
\frac{\theta_2(x_Q\! +\! i \gamma \frac{n_0}{K^2}, 
e^{-2 \gamma/K^2})\theta_2(x_\Pi, e^{-2 \gamma K^2})
+ \theta_3(x_Q \!+\! i \gamma \frac{n_0}{K^2}, 
e^{-2 \gamma/K^2})\theta_3(x_\Pi, e^{-2 \gamma K^2})}{
\theta_2( i \gamma \frac{n_0}{K^2}, 
e^{-2 \gamma/K^2})\theta_2(0, e^{-2 \gamma K^2}) 
+ \theta_3(i \gamma \frac{n_0}{K^2},
e^{-2 \gamma/K^2})\theta_3(0, e^{- 2 \gamma K^2})}
\end{equation}
where $\gamma \equiv \frac{\pi v \beta}{\ell}$ is a measure of the energy
gap compared to the temperature scale, and we have defined $x_Q \equiv
 \frac{\pi}{\ell}(x + vt/K^2)$ and $x_\Pi \equiv \frac{\pi}{\ell}(x + vt K^2)$.
The complete chiral Green's function therefore reduces to a compact 
expression
\begin{equation} 
G(x,t) \propto  B(x,t) 
\left[ \frac{\theta_1(\pi\frac{x+vt}{\ell},
e^{-\pi v \beta/\ell})}{\theta_1(-i\frac{\pi \alpha}{\ell},  
e^{-\pi v \beta/\ell})} \right]^{-(K + \frac{1}{K})^2/4}
\left[\frac{\theta_1(\pi\frac{vt-x}{\ell},
e^{-\pi v \beta/\ell})}{\theta_1(-i\frac{\pi \alpha}{\ell},  
e^{-\pi v \beta/\ell})} \right]^{-(K - \frac{1}{K})^2/4}. \label{result}
\end{equation}
We can immediately verify that this expression
is anti-periodic under translation $t \to t +i\beta$ and periodic 
under translation $x \to x+\ell$ up
to a phase of $e^{ik_F \ell}$ as it should be.

If the temperature is much smaller than the energy gap of the system, 
i.e.~$\gamma=\frac{\pi v \beta}{\ell} \to \infty$, we can 
use the limits of the theta function\cite{GR} 
as $q = e^{-\gamma} \to 0$ to find  
\begin{equation}
G(x,t) \to e^{i(n_0\,{\rm mod} \,
2) x_Q} \left[ \frac{\ell}{\pi} \sin(\pi\frac{x+vt}{\ell})
\right]^{-(K + \frac{1}{K})^2/4}
\left[\frac{\ell}{\pi} \sin(\pi\frac{vt-x}{\ell})
\right]^{-(K - \frac{1}{K})^2/4} \label{finiteL}
\end{equation}
which is the expected finite length result\cite{Voit}. 
For special values of $n_0$
the overall phase may have a different x-dependence, which will not be 
discussed here in detail.  

Likewise, we can explore the limit of a large system size compared to the
temperature scale, i.e.~$\gamma \to 0$.  
We use the Poisson summation formula to find
$\theta_1(z,e^{-\gamma}) \approx 2 \sqrt{\frac{\pi}{\gamma}} e^{-z^2/\gamma}
e^{-\pi^2/4 \gamma} \sinh \frac{\pi z}{\gamma}$ and 
$\theta_2(z,e^{-\gamma}) \approx \theta_3(z,e^{-\gamma}) \approx
\sqrt{\frac{\pi}{\gamma}} e^{-z^2/\gamma}$   as $e^{-\gamma} \to 1$
so that the correct finite temperature result\cite{frahm2} is reproduced
\begin{equation}
G(x,t) \to \left[ \frac{v \beta}{\pi}\sinh(\pi\frac{x+vt}{v \beta})
\right]^{-(K + \frac{1}{K})^2/4} 
\left[\frac{v\beta}{\pi}\sinh(\pi\frac{vt-x}{v\beta})
\right]^{-(K - \frac{1}{K})^2/4}. \label{finiteT}
\end{equation}
Taking the limit $l \to \infty$ in equation (\ref{finiteL}) or 
$v \beta \to \infty$ in equation (\ref{finiteT}) recovers the
well known universal power-laws. 

One interesting aspect of equations (\ref{zeromodes}) and (\ref{result}) is
the dependence on the variable $n_0 = \frac{k_F \ell}{\pi} + 1$. 
The  Green's function (\ref{result}) is invariant under the increase 
of $n_0$ by 2, i.e.~periodic in $k_F$ with period $\frac{2 \pi}{\ell}$.
Hence, as we change the Fermi energy (e.g.~by applying a gate voltage), the 
spectral properties of the system are changed periodically.  These
are Coulomb blockade oscillations\cite{fazio,Krive}, which come from
the fact that the system has a regular energy-level spacing. 
In the case of Luttinger liquids this level spacing has been derived 
from a complete solution of 
a quantum system which takes all interactions into account and cannot be 
obtained from a geometrical analysis of the capacitance as in the usual 
quantum-dot systems.  As possible experimental setups to test this 
periodicity we can imagine electron tunneling through 
a quasi one-dimensional wire or ring of mesoscopic size.  Such 
wires can be fabricated by etching, gating\cite{Tarucha} or 
by cleaved edge overgrowth\cite{yacoby}.  A small gated quantum hall
bar on the other hand should exhibit {\it chiral} Luttinger liquid 
behavior which can be examined with the same formalism, 
except that in that case the rescaling parameter $K$ is fixed
by the filling fraction in the bulk of the quantum Hall bar\cite{Wen}.
The experimental 
situation is developing quickly and most recently carbon nanonubes have been 
produced which were seen to exhibit the charactersitcs of one dimensional
wires\cite{nano}.

\begin{figure}
\begin{center}
\mbox{\epsfxsize=6.3in \epsfbox{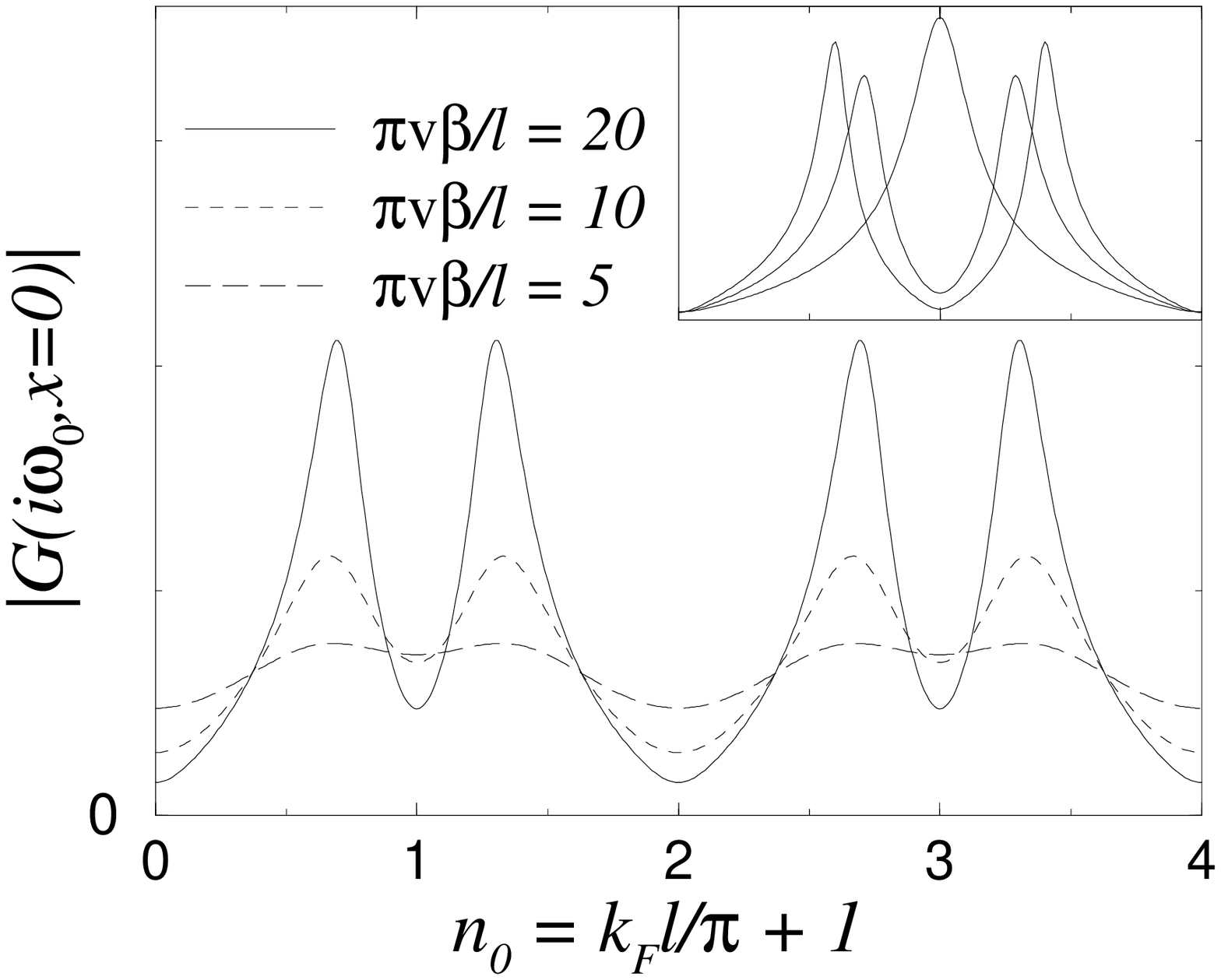}}
\end{center}
\caption{The absolute value of the Green's function at the 
Matsubara frequency $i \omega_0 = i \pi/\beta$ 
as a function of the Fermi wave-vector $k_F$ for different  
temperatures and $K^2=0.7$. 
The periodic resonances in the spectral properties
are the so-called Coulomb blockade oscillations. The inset shows the 
curves for different interaction parameters $K^2= 0.5,\ 0.7,\ 1$ at
$\frac{\pi v \beta}{l}=30$.} 
\label{oscillations}
\end{figure}

As an example we consider the Green's function at the lowest Matsubara 
frequency $i \omega_0 = i \pi/\beta$ for spin-less Fermions on a ring.
This quantity only gives an indirect indication of experimental tunneling 
resonances, but it is a simple illustration of the properties of
equation (\ref{result}) as a function of temperature and interaction.  
The absolute value of the Matsubara Green's function has been plotted
in figure (\ref{oscillations}) as a function of the Fermi wave-vector
in arbitrary units.  As the Fermi-level is changed, the system 
shows periodic resonances which become more pronounced
as the temperature is lowered.  These peaks occur
at special values of $n_0 = \frac{1+K^4}{2},\ \frac{3-K^4}{2}$ 
where the even and odd 
sectors of $n$ and $m$ in equation (\ref{quanta}) give the same degenerate
ground state.  Hence, the spacing of the resonances can give a direct 
experimental measure of the interaction strength $K$\cite{kinaret}.
As shown in the inset of figure (\ref{oscillations}) a non-interacting
system ($K=1$)
has only one central peak in the range $0 \leq n_0 < 2$, while an interacting
system has two split resonances.
 The physical interpretation of the resonances
is that tunneling of an extra electron into the system does not require
any additional energy at the degenerate values of $n_0$ so that such 
processes are enhanced.
Recently, much progress has also been made with 
Bethe ansatz methods\cite{frahm}, which showed evidence of
Coulomb-blockade-like oscillations 
in a strongly interacting system, but in contrast to statements made
in that paper, field theoretical methods also 
predict the desired Coulomb blockade oscillations\cite{fazio,Krive}.
Using our result (\ref{result}) it is now possible to determine
the complete spectral properties,
including the dependence of the Coulomb blockade oscillations on
interactions and on temperature for any of the systems mentioned above.

So far we have taken the fermion field to be spinless, but the
results apply equally well to electrons with spin because the well 
known spin-charge separation allows the  
same formalism to be used for the spin and the 
charge excitations separately\cite{Voit}. 
It is also well understood how to incorporate open boundary condition within
the same formalism\cite{boundary,mattsson}.

To test our result, we will now consider
one of the simplest models of spinless interacting fermions described
by the Hamiltonian
\begin{equation}
H = - t \sum_x \left[\psi^\dagger(x) \psi^{}(x+1) + h.c.\right] + U  \sum_x
\left[n(x)-\frac{1}{2}\right] \left[n(x+1)-\frac{1}{2}\right], \label{lattice}
\end{equation}
where $n = \psi^\dagger \psi^{}$, and the lattice constant has been
 set to unity.  
In this case the coupling constants are given by $g_1 =  g_2/4 = U$. 
We have chosen the system to be at half-filling 
$k_F = \pi/2$ where the model is equivalent to the xxz spin chain. In this
case the umklapp process is allowed, but irrelevant, and the 
rescaling parameter $K$ and the velocity $v$ can actually be determined 
exactly by comparison with Bethe ansatz results\cite{bethe}, 
i.e.~to all orders in the coupling constants.  For our
numerical simulations we have taken $t = U$ which 
corresponds to $K^2 = \frac{3}{4}$ and $v =  t \frac{3 \sqrt{3}}{2}$. 
In particular, we consider the equal time
density-density correlation function $\left<n(x)n(0)\right>$. 
At half filling the density-density correlation
function acquires an alternating part which dominates the uniform
contribution and can be expressed in terms of chiral fermions

\begin{equation}
\left<n(x) n(0)\right>_{\rm alt} 
\propto e^{2ik_F x} \left< \psi_L^\dagger(x) \psi_R^{}(x)
\psi_R^\dagger(0) \psi_L^{}(0) \right>.
\end{equation}
Following the analogous steps from equation 
(\ref{GF}) to equation (\ref{result}) we find
\begin{equation}
\left<n(x) n(0)\right>_{\rm alt}  \propto   (-1)^x 
\left| \frac{\theta_1(\frac{\pi x}{\ell},
e^{-\gamma})}{\theta_1(\frac{i\pi 
\alpha}{\ell},  e^{-\gamma})}
\right|^{-3/2} \label{density} 
\frac{\theta_2(\frac{ 2\pi x}{\ell}, e^{-8\gamma/3})\theta_3(0, e^{-3 \gamma/2})
+\theta_3(\frac{ 2\pi x}{\ell}, e^{-8\gamma/3})\theta_2(0, e^{-3 \gamma/2})}{
\theta_2(0, e^{-8\gamma/3})\theta_3(0, e^{-3 \gamma/2})
+\theta_3(0, e^{-8\gamma/3})\theta_2(0, e^{-3 \gamma/2}) }. 
\end{equation}
Here we took $n_0$ to be odd, i.e.~$\ell$ divisible by four,
but other cases have similar expressions.

\begin{figure}
\begin{center}
\mbox{\epsfxsize=6.3in \epsfbox{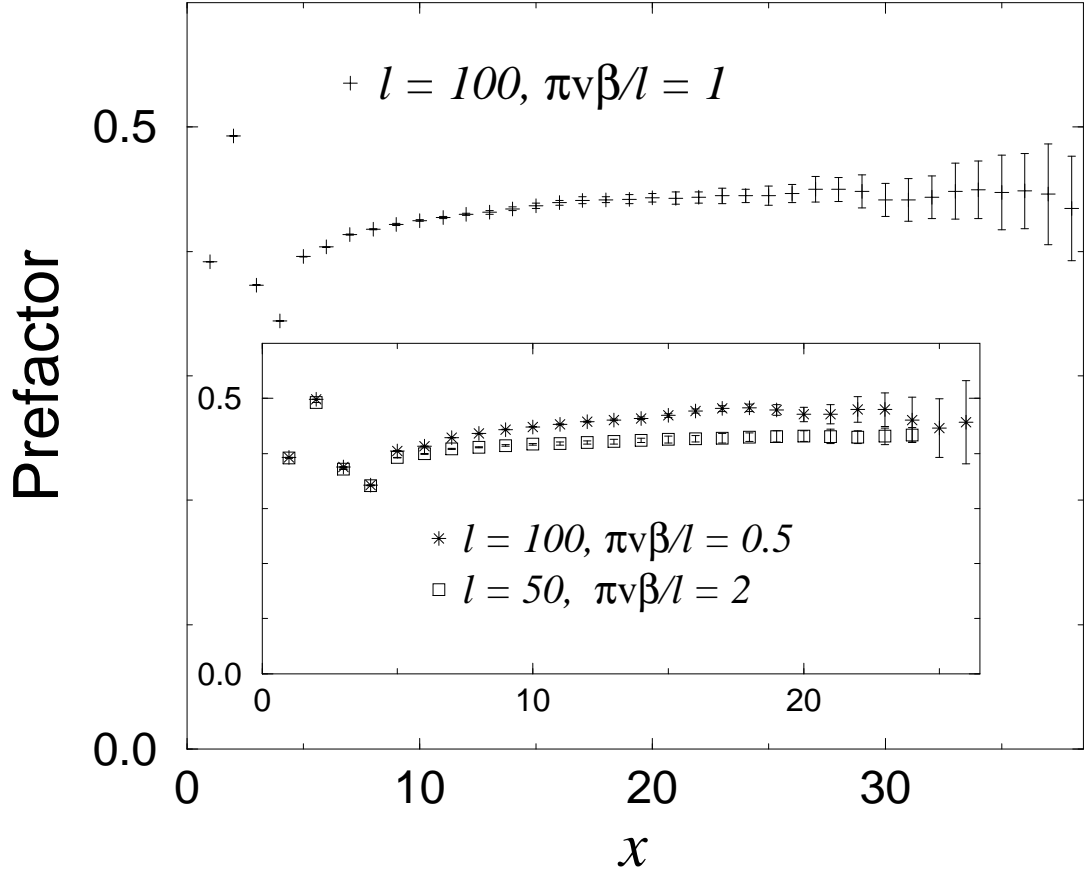}}
\end{center}
\caption{The prefactor of equation (\protect{\ref{density}}), which is 
determined by dividing the alternating part of the density-density
correlations from numerical simulations by the expressions of equation
(\protect{\ref{density}}) for each correlation length  $x$ separately.
As predicted, the prefactor is asymptotically constant in all cases.}
\label{prefactor}
\end{figure}

For comparison we performed numerical
Monte Carlo simulations using the Hamiltonian
(\ref{lattice}) and extracted the alternating part of the density-density 
correlation function to determine the unknown proportionality constant
in equation (\ref{density}).
This prefactor should be independent of $x$ up to corrections for
smaller distances, which come from irrelevant terms 
in the Hamiltonian that can be neglected in the long-distance 
limit $x \gg \alpha$.
As shown in figure (\ref{prefactor}) 
the prefactor is indeed constant within the error-bars
of the Monte Carlo simulations for all system sizes $\ell$ 
and temperatures. This serves as a very sensitive test of our 
theoretical predictions since
the expression (\ref{density}) varies over
several orders of magnitude over the range shown, so that even slight
changes from the predicted expression (\ref{density})
would have resulted in huge
deviations from a constant for larger values of $x$.  
Using the previously known approximate expressions 
(\ref{finiteL}) or (\ref{finiteT}) is not sufficient in this case.

We therefore conclude that our field 
theoretical calculations make very accurate estimates
for the correlation functions of almost any one-dimensional interacting
system at arbitrary temperature and system size (as long as the
cutoff $\alpha$ is small compared to the corresponding scales).
The results can be used to determine the complete spectral behavior
of such systems, including Coulomb blockade oscillations as
a function of both interaction and temperature.  Monte Carlo simulations
serve as accurate ``experiments'', which confirm the results. 

\begin{acknowledgements}
The authors would like to thank Henrik Johannesson 
 for helpful discussions. This research was supported in part 
by the Swedish Natural Science Research Council.
\end{acknowledgements}

\end{document}